\documentclass[english,aps,prl,reprint,superscriptaddress]{revtex4-1}
\usepackage[T1]{fontenc}
\usepackage[utf8]{inputenc}
\usepackage{float}
\usepackage{units}
\usepackage{amsmath}
\usepackage{graphicx}

\makeatletter
\usepackage{babel}

\makeatother

\usepackage{babel}
\begin{document}

\title{Reversibility of granular rotations and translations}

\author{Anton Peshkov}

\affiliation{IREAP, University of Maryland, College Park, Maryland, USA}

\author{Michelle Girvan}

\affiliation{Departments of Physics, IPST and IREAP, University of Maryland, College
Park, Maryland, USA}

\author{Derek C. Richardson}

\affiliation{Department of Astronomy, University of Maryland, College Park, Maryland,
USA}

\author{Wolfgang Losert}

\affiliation{Departments of Physics, IPST and IREAP, University of Maryland, College
Park, Maryland, USA}
\begin{abstract}
We analyze reversibility of displacements and rotations of spherical
grains in three-dimensional compression experiments. Using transparent
acrylic beads with cylindrical holes and index matching techniques,
we are not only capable of tracking displacements but also, for the
first time, analyzing reversibility of rotations. We observe that
for moderate compression amplitudes, up to one bead diameter, the
translational displacements of the beads after each cycle become mostly
reversible after an initial transient. By contrast, granular rotations
are largely irreversible. We find a weak correlation between translational
and rotational displacements, indicating that rotational reversibility
depends on more subtle changes in contact distributions and contact
forces between grains compared with displacement reversibility. 3D
rotations in dense granular systems are particularly important, since
frictional losses associated with rotations are the dominant mechanism
for energy dissipation. As such our work provides a first step toward
a thorough study of rotations and tangential forces to understand
the granular dynamics in dense systems.
\end{abstract}
\maketitle

\section{Introduction}

Flows of granular matter are an important subject of study in many
fields, including geology (where they are relevant for avalanches
and earthquakes), engineering (where they play a role in industrial
processes and construction), and astronomy (where they affect the
formation of asteroids and planets). In these contexts, it is particularly
important to understand how the granular material transitions from
a jammed state to a state of flow with irreversible rearrangements.
A common driver of granular flow are cyclic perturbations, which can
be caused by a wide range of driving forces, such as vibrating apparatus
(construction), periodic loads (roads and rails), earthquakes, cyclic
gravitational fields (tidal forces), or thermal expansion and contraction
due to day/night temperature variations (on planets and asteroids).
Beside the question of flows, the cycling compression or shearing
of the granular system can change the properties of the packing, leading
to compaction \cite{HECKE:2010:ID749} and ultimately crystallization
in some cases \cite{RIETZ:2018:ID833,PANAITESCU:2012:ID748}.

Most past numerical and experimental work in this area has concentrated
on the translational motion of particles and the associated force
fields. However, granular systems are frictional and often aspherical,
which implies the presence of rotations of the grains. Rotations and
torques can play an important role in mesoscopic granular motions,
as has been shown in two-dimensional investigations \cite{TORDESILLAS:2010:ID746}.
Thus bulk granular mechanics is affected by both force networks and
torques, though rotations and torques are often ignored, with the
notable exception of the Cosserat continuum elastic models \cite{SRINIVASA:2002:ID755}.

Furthermore, rotations are likely more important in a frictional three
dimensional system, where friction will increase the number of constraints
from 3 to 6. Compare this to a model two dimensional system with an
increase from 2 to 3. Frictional, rotating, systems can become isostatic
with four contacts per particle, long before the jamming point at
6 contacts per particle \cite{HECKE:2010:ID749} and are therefore
over-constrained. The distribution of frictional forces is dependent
on the history of contacts \cite{KASAHARA:2004:ID750}, i.e., the
method of preparation of the system. Periodic compression will alter
the contact distribution and could lead to a different state of the
system, called a random loose packing, in contrast to the random close
packing with six contacts per particle that we would find for a frictionless
system. Thus understanding rotations is important to elucidate the
nature of different jammed states of frictional spherical (or aspherical)
particles as well as their transition to flow.

Despite the importance of rotations for granular systems, only a handful
of experimental investigations have studied them \cite{MATSUSHIMA:2003:ID836,ZHANG:2010:ID835},
especially in three dimensions \cite{WENZL:2013:ID840,GUILLARD:2017:ID662,KOU:2017:ID758,KOU:2018:ID757,HALL:2010}.
In this study of reversibility we build on our first experimental
measurements of particle-scale rotations in a three-dimensional granular
flow \cite{HARRINGTON:2014:ID751}, and systematically measure both
rotations and translations under cyclic forcing. To focus our investigation
on irreversibilities associated with rotational motion, we study small
enough driving amplitudes so that the translational motion of the
system is reversible. We had previously shown that in this small forcing
regime, convective flows and segregation—two hallmarks of bulk granular
flow—are also suppressed \cite{HARRINGTON:2013:ID668}.

\section{Experimental setup}

\begin{figure*}
\includegraphics[width=1\textwidth]{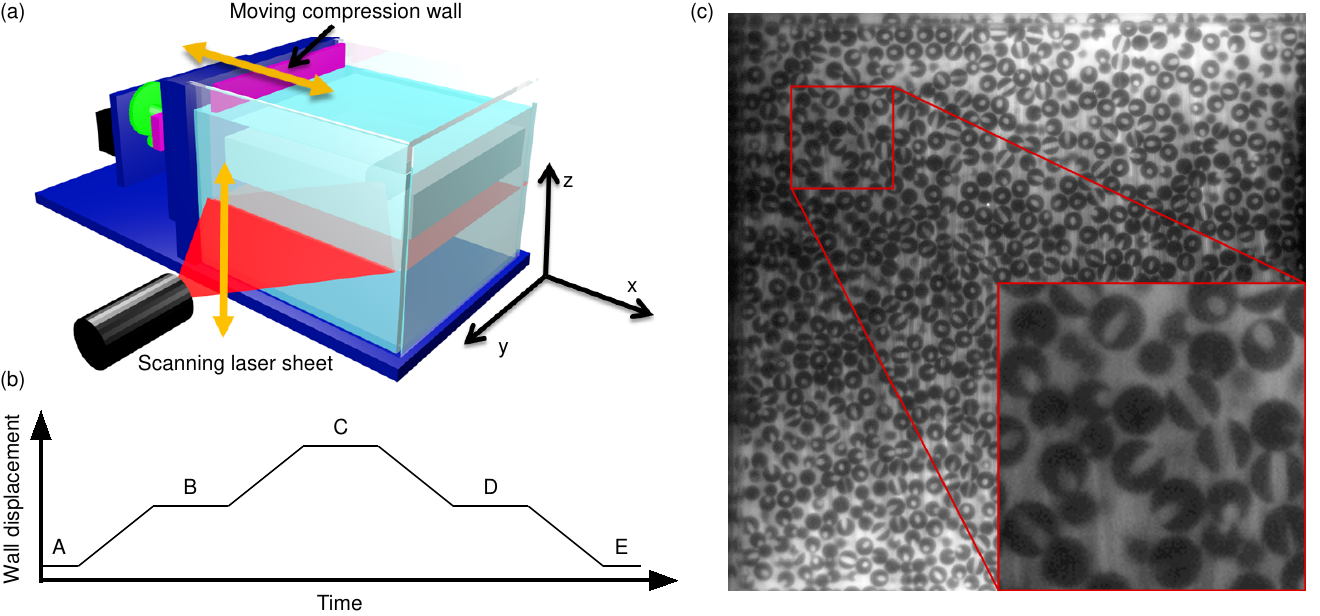}

\caption{\label{fig:Experimental setup}(a) Schematics of the setup. (b) Schematics
of the 5 positions of the wall at which 3D scans are performed. Note
that position E of the current cycle correspond to position A of the
next cycle. (c) One example image of one hole beads - typically a
sequence of 500 images is captured in each 3D scan. A whole scanning
sequence can be seen in supplemental movies.}
\end{figure*}

The experimental setup, sketched in figure \ref{fig:Experimental setup}(a),
is composed of a container with transparent walls filled with transparent
acrylic beads. The container is 15.24 cm wide ($y$ direction) and
approximately 15 cm long ($x$ direction). The back wall is displaced
by a motor to compress the whole system. The amplitude of compression
was varied between 0.5\,\% and 2.5\,\% of the container length.
Note that we can not go beyond 3\% with our setup, as it corresponds
to the diameter of a single bead. Beyond this limit, the beads will
jam between the weight placed on the beads and the moving wall. At
these amplitudes, our experimental system is still below the convection
regime, which will invariably appear at higher amplitudes \cite{ROYER:2015:ID759}.
The compression speed was 0.05~mm~s$^{-1}$ for all amplitudes,
i.e. a constant shear rate. We verified that this speed is slow enough
to expect granular rearrangements that resemble dry systems \cite{DIJKSMAN:2012:ID663}.
All the experiments contained between 500 and 1000 cycles of compression,
with between 4 and 16 three-dimensional images captured during each
cycle. All figures in the manuscript are presented for the case of
2.5\,\% compression, unless noted otherwise.

The container is filled with 20,000 transparent acrylic beads of radius
$R=2.5\,mm$, forming a height ($z$ direction) of approximately twenty
layers. Each bead has a $1.4\,mm$ diameter cylindrical cavity running
through it, allowing us to detect the orientation of the bead. A $1.5\,kg$
transparent acrylic weight is placed on the beads to maintain a constant
pressure. The beads and the weight are submerged in an index-matched
solution of Triton X-100, which additionally includes a fluorescent
Nile blue dye as well as 0.05\,\% of hydrochloric acid (38\,\% concentration)
. The whole system is illuminated with two 660 nm, which approximately
corresponds to the maximum excitation fluorescence frequency of Nile
blue, scanning sheet lasers from both sides of the container.

We should note that a single hole in the grains does not allow us
to detect rotations when the axis of rotations is along the hole of
the bead. Only rotation components perpendicular to the hole can be
resolved. Thus, our current experimental setup captures only a part
of the rotation amplitude. However, in this work we focus on the statistical
analysis of amplitudes of rotations, not taking into account such
information as the axis of rotation or its correlation with other
quantities. Given the small compression amplitude, the beads rotate
less than $\nicefrac{\pi}{2}$ between scans, preventing any degeneracy
in the detection of rotations.

Typically four full scans of the system are performed per compression
cycle, as shown in figure \ref{fig:Experimental setup}.b) : a scan
before compression (time A), a scan at half-compression (B), a scan
at full compression (C), a scan at half-decompression (D) and a scan
at full decompression (E). The first and last 10 cycles are recorded
with 16 scans per cycle. Most of the presented results are an average
over the last 10 cycles of the experiment.

Positions and rotations are then extracted and tracked \cite{HARRINGTON:2014:ID751}.
To directly compare the rotations and translations in the same units,
we multiply all rotation angles by the radius of the bead. In the
case of non-slip rolling of a bead on a flat surface, this will make
the values of rotational and translation displacements identical.
We have estimated the standard error on particle position detection
to be $20\,\mu m$ and the standard error on angle detection to be
$0.0345\,rad$ ($\sim2^{\circ}$) or $86\,\mu m$.

Crystallization occurs near the borders of the experimental cell,
for the statistical analysis we omit all particles in this layers
as well as in the bottom four layers, where optical aberrations lead
to larger detection errors and all the particles whose motion during
the compression cycle is less than the standard error.

\section{Validation of the experimental approach}

The hole in the middle of the bead produce two ``cavities'' on the
bead surface in which other beads can ``fall''. It is thus necessary
to verify if the beads have a preference to form a contact at the
cavities and if yes, to verify that it does not significatively change
the results that we obtain. 

\begin{figure}
\includegraphics[width=1\columnwidth]{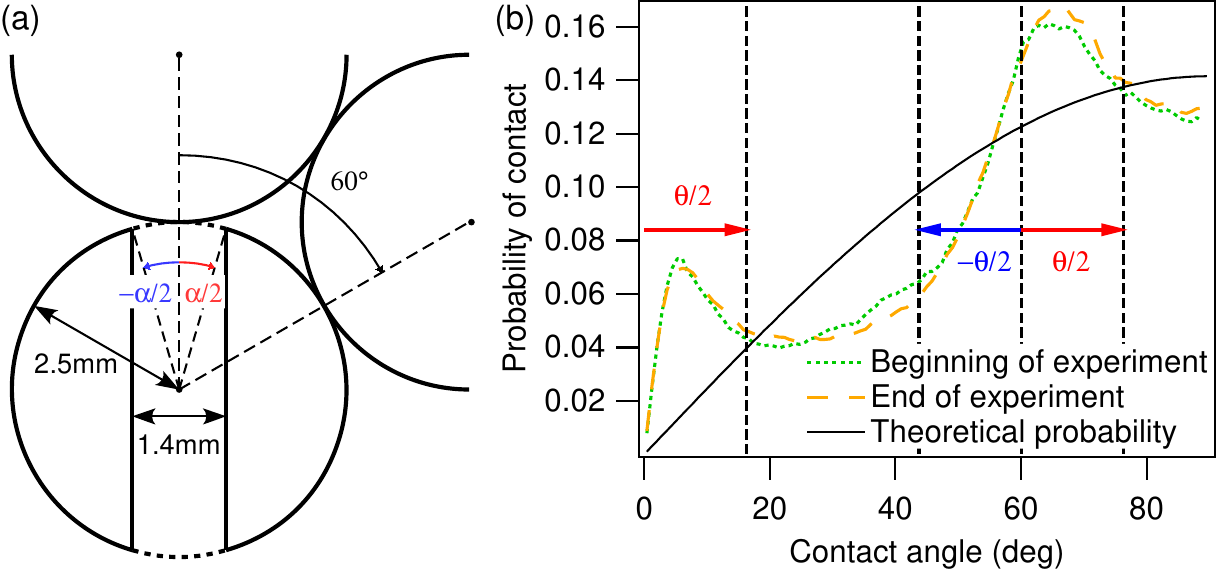}

\caption{\label{fig:Hole contact probability}a) Sketch of the contact through
a hole. Note that an increase in the probability of contacting through
a hole will also increas the probability of a contact hapening at
$60^{\circ}$ from the hole axis. b) Experimental probability of a
contact point being at a certain angle to the hole axis. Green dots
indicate the begining of the experiment, yellow dash the end of experiment
and black line the theoretical predicted value for an equal probability
contact point. Results for 2.5\% compression amplitude.}
\end{figure}

Let us call $\alpha$ the angular oppening of the hole as shown on
figure \ref{fig:Hole contact probability} a). A straitforward calculation
will give 
\[
\alpha=2\sin^{-1}\left(\frac{d}{D}\right)
\]
where $D=5\thinspace mm$ is the diameter of the bead and $d=1.4\thinspace mm$
is the diameter of the hole. Given an axis passing through the hole,
the cumulative probability that a contact point will be located within
an angle $\beta\in$$\left[0:\nicefrac{\pi}{2}\right]$ of this axis
is given by the surface area located within this angle. We can then
easily show that this probability to be located within an angle $\beta$
from any of the two holes is
\begin{align*}
P\left(\beta\right) & =1-\cos\left(\beta\right)\\
 & =1-\sqrt{1-\sin^{2}\left(\beta\right)}
\end{align*}
.

If we define a contact through a hole as any contact whose contact
point is located within an angle $\nicefrac{\alpha}{2}$ of the hole
axis, then the theoretical probability of contacting through the hole
is given by
\begin{align*}
P\left(\frac{\alpha}{2}\right) & =1-\sqrt{1-\left(\frac{d}{D}\right)^{2}}\\
 & =4\thinspace\%
\end{align*}
. To check this prediction we compute the probability of a contact
hapening at a particular angle from the axis from our experimental
data. To detect a contact we consider all beads located within a cuttof
distance $r_{c}$ as being in contact. The value of $r_{c}$ is determinde
as described in the part \ref{sec:Contact-dynamics} of the article.
We compute this probability both at the begining and the end of the
experiment. The result is presented on figure \ref{fig:Hole contact probability}
b). We can see that the actual probability of contacting through a
hole is higher than the theoretically predicted value indicating that
indeed the particles tend to ``fall'' inside the hole. The experimental
cummulative probability of contacting through the hole is $\approx12\thinspace\%$.
Note that the increased probabily of a contact hapening through the
hole, also increase the probability of a contact happening at $60^{\circ}$
from the hole as explained on figure \ref{fig:Hole contact probability}
a). This results are in good agreement with the predictions made for
``shpero-cylindrical'' particles \cite{MARSCHALL:2019:ID863}. However
we can see that the probability of contacting through a hole is unchanged
at the begining of the experiment as well as at the end of the experiment.
This means that there is no any tendency for the ``crystalization''
of hole contacts.

\begin{figure}
\includegraphics[width=1\columnwidth]{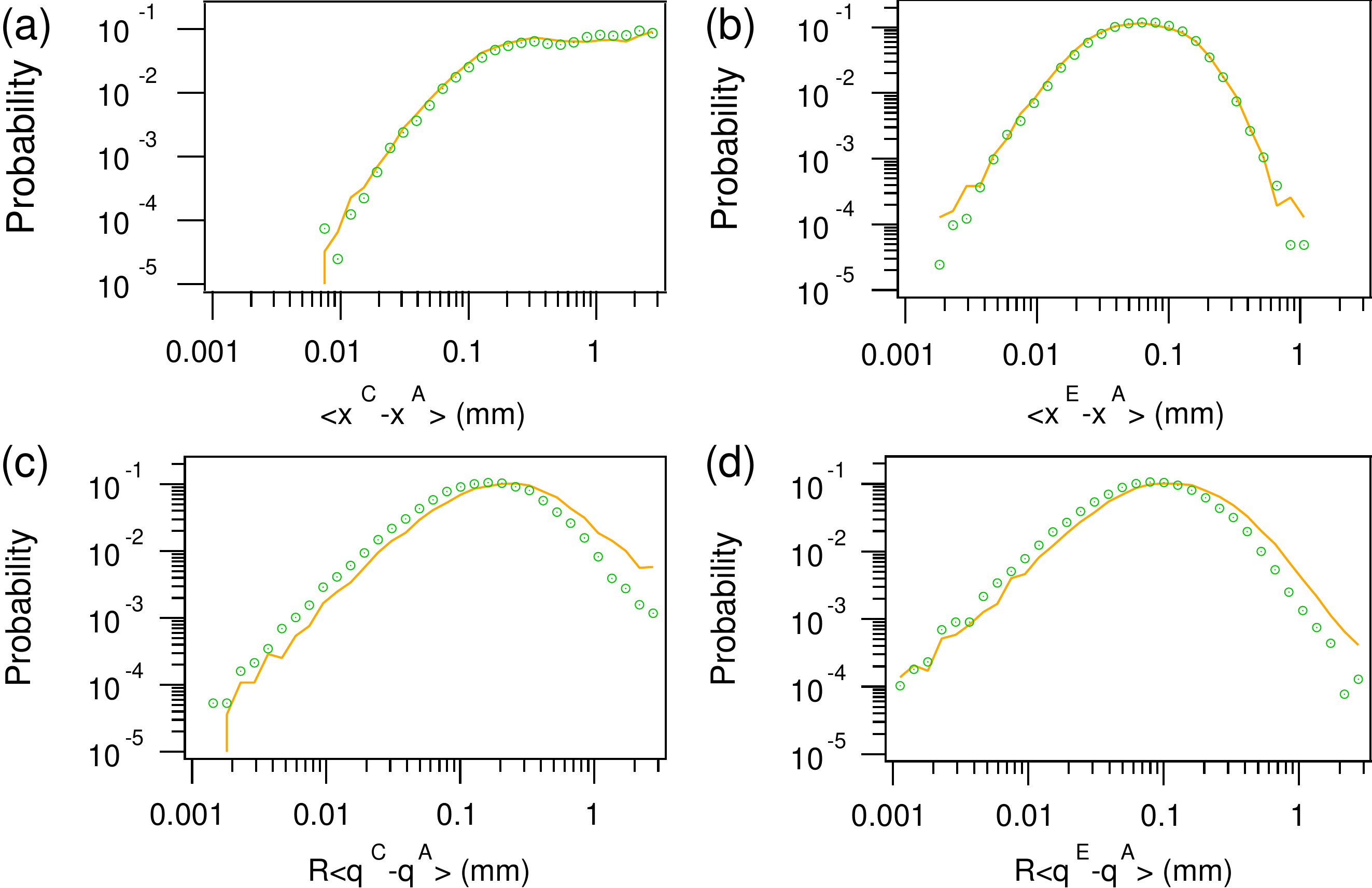}

\caption{\label{fig:Probability of displacement}Probability of translational
( a and b) and rotational ( c and d) displacements between the begining
and middle ( a and c) and end ( b and d) of cycle. Yellow line indicate
particles that do not contact through their hole at the begining of
the cycle, while green circles indicate particles that do contact
through the hole. Displacement units in mm. Results for 2.5\% compression
amplitude.}
\end{figure}

We can ask the question if the particle contacting through a hole
will see their dynamics impacted as to the particles which do not
contact through the hole. For this end we measure the probability
of translational and rotational displacements at times $C$ and $E$
as compared to the begining of the cycle. We measure these quantities
separately for particles that do contact through a hole at the begining
of the cycle and particles that do not. The results are presented
on figure \ref{fig:Probability of displacement}. We can see that
the translations are unaffected by the type of contact. Only rotations
suffer a small decrease in their amplitude while mantaining the overall
shape of the distribution. Note that this decrease in rotational mobility
emphasis even more our funding on the ireversibility of the rotational
dynamics.

\section{Translational and rotational displacements}

We denote by $\vec{x}_{i}^{T}$ the position of particle $i$ at the
temporal position $T$ (one of A–E) in the cycle and by $\hat{q}_{i}^{T}$
its orientation expressed as a unit vector. We introduce the relative
(as compared to the beginning of the cycle) displacements of particles
at time $T$: the translational $\mu_{i}^{T}=\left\Vert \vec{x}_{i}^{T}-\vec{x}_{i}^{A}\right\Vert $
and the rotational $\theta_{i}^{T}=\sin^{-1}\left(\left\Vert \hat{q}_{i}^{T}\times\hat{q}_{i}^{A}\right\Vert \right)$,
as well as their averages: $\left\langle \mu^{T}\right\rangle =\frac{1}{N}\sum_{i=1}^{N}\mu_{i}^{T}$
and $\left\langle \theta^{T}\right\rangle =\frac{1}{N}\sum_{i=1}^{N}\theta_{i}^{T}$.

\begin{figure}
\includegraphics[width=1\columnwidth]{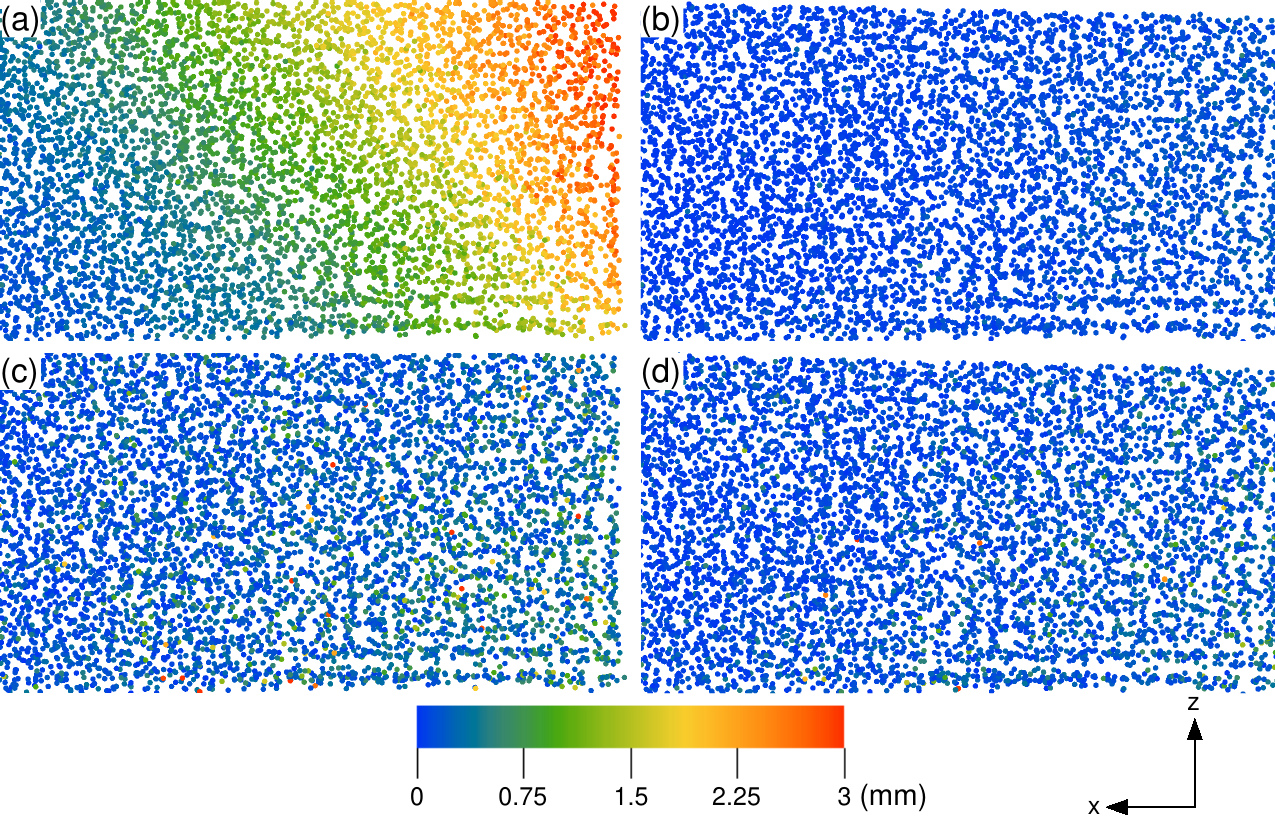}

\caption{\label{fig:Displacement-projections}Two dimensional projections on
the x-z plane of bead translational (a and b) and rotational (c and
d) displacements between the begining and middle (a and c) and between
the begining and end (b and d) of a cycle. The color indicate the
amplitude of displacement. The compression wall is located to the
right of the snapshots. All units in mm.}

\end{figure}

Figure \ref{fig:Displacement-projections} a) shows a snapshot of
typical translational displacements of the particles at time $C$.
The gradient angle matches the symmetry of the system, with a compressing
right wall and movable top wall kept at constant pressure. In contrast,
the rotations, represented on the \ref{fig:Displacement-projections}
c), do not present any apparent shear zones, with a larger concentration
of rotations near the bottom of the cell, suggesting an absence of
direct correlation between translations and rotations. This is corroborated
by statistical analysis as presented below. Figure \ref{fig:Displacement-projections}
b) and d) shows snapshots at the end of the cycle, at time $E$, of
translations and rotations respectively. While most of the translational
displacement of the particles is reversed, this is not the case for
all rotations as discussed further.

Note that due to the nature of our experiments, a small but measurable
compaction (following a power law) is found as expected \cite{BANDI:2018:ID834}
in our system. While the particle positions are not perfectly reversible,
we focus on small amplitudes where particles retain their neighbors,
and where (as shown in \cite{HARRINGTON:2013:ID668}) convection and
segregation are suppressed. In this regime, where the position of
most particles with respect to their neighbors is reversible, we investigate
rotations, as well as their evolution with cycle number.

\begin{figure}
\includegraphics[width=1\columnwidth]{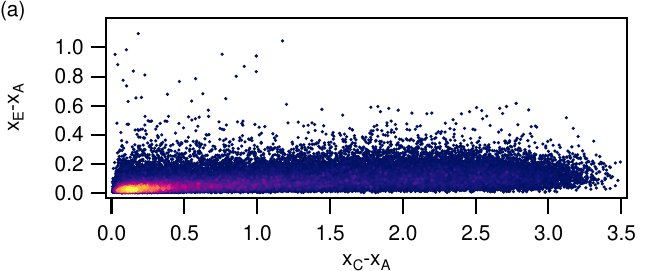}

\includegraphics[width=1\columnwidth]{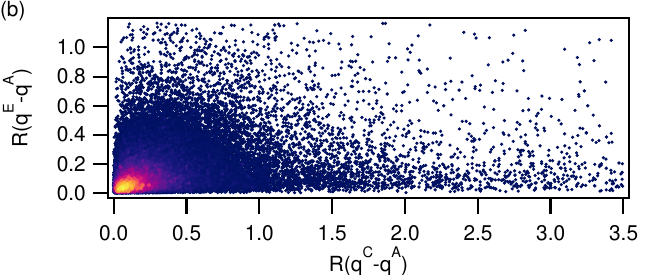}

\caption{\label{fig:Displacement scatter plots}(a) Scatter plot of translations
between the middle of the cycle and the end of cycle. (b) Same for
rotations. The color indicates the beads density at the particular
location. All units in mm. Results for 2.5\% compression amplitude.}
\end{figure}

To examine the irreversibility of the granular motion we can study
the translational and rotational displacements of particles at the
end of the cycle: $\mu_{i}^{E}$ and $\theta_{i}^{E}$. As can be
seen in scatter plots of translations and rotations (figure \ref{fig:Displacement scatter plots}),
comparing the displacements of each particle at time $C$ to the displacements
at time $E$, translations are reversible as expected, while rotations
are mostly irreversible with a lot of particles not returning to their
initial orientation.

\begin{figure}
\includegraphics[width=1\columnwidth]{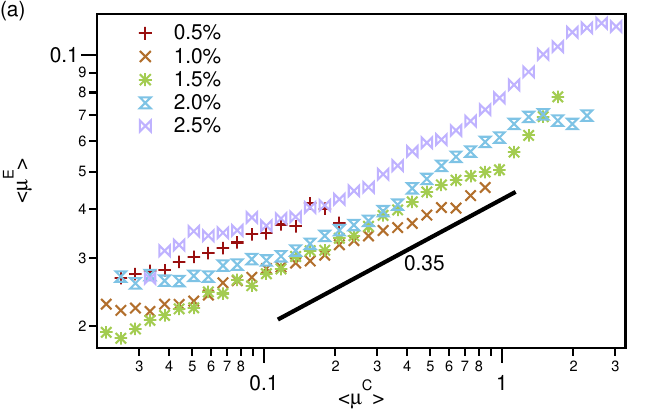}

\includegraphics[width=1\columnwidth]{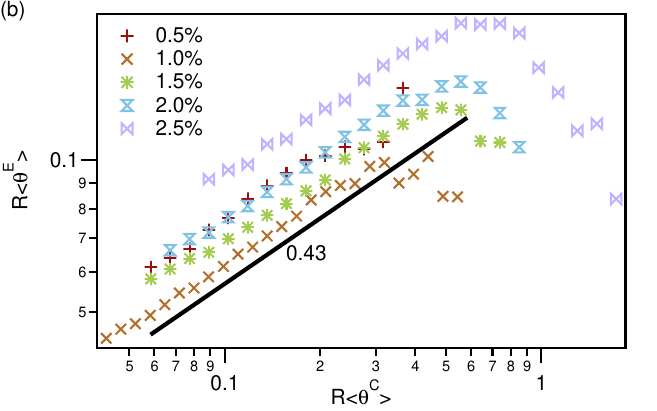}

\caption{\label{fig:Mobilties-comparison}The expected end of cycle translational
(a) and rotational (b) displacements as compared with the mid-cycle
displacements, for different compression amplitudes. The solid lines
illustrate power laws with exponents of 0.35 and 0.43. All units in
mm.}
\end{figure}

To obtain a better statistical understanding of the phenomena, we
compare the expected end-of-cycle displacements $\mu^{E}$ and $\theta^{E}$
for given values of mid-cycle displacements $\mu^{C}$ and $\theta^{C}$
(figure \ref{fig:Mobilties-comparison}). These are well characterized
by power-law behaviors with approximate exponents of $\approx0.35$
for translations and $\approx0.43$ for rotations at moderate displacements.
As seen in the figure, these exponents do not depend on the amplitude
of compression of our system, indicating a universal behavior for
the amplitudes we investigated. Note that such a behaviour could indicate
an existence of strong temporal correlations for the translations
and rotations\footnote{A. Peshkov et al., under preparation.}.

\begin{figure}[H]
\includegraphics[width=1\columnwidth]{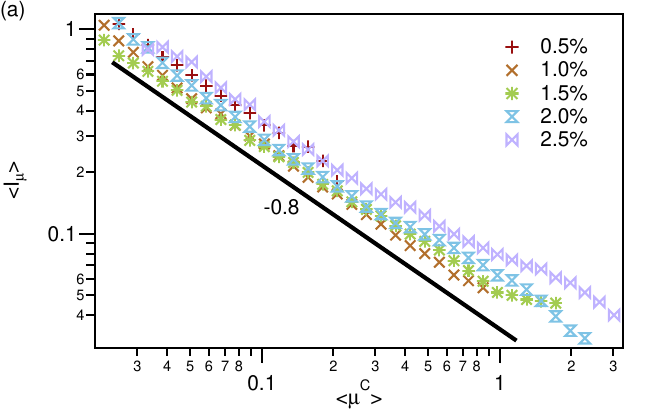}

\includegraphics[width=1\columnwidth]{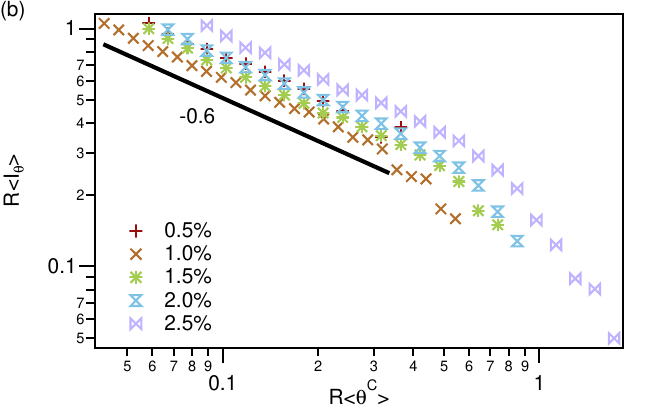}

\caption{\label{fig:Irreversibility}Relative irreversibility parameters $I_{\mu}$
(a) and $I_{\theta}$(b) as a function of the mid-cycle mobility $\mu^{C}$
and $\theta^{C}$ for different compression amplitudes. All units
in mm. }
\end{figure}

At very large values of translation displacement the exponent seems
to increase, probably corresponding to a transition to diffusive behavior
when beads move so much that they can escape their ``cage''. Both
the translational and rotational exponents are smaller than 1, which
implies that the relative reversibility improves for bigger displacements.
We introduce relative irreversibility parameters $I_{\mu,i}=\nicefrac{\mu_{i}^{E}}{\mu_{i}^{C}}$
for translations and $I_{\theta,i}=\nicefrac{\theta_{i}^{E}}{\theta_{i}^{C}}$
for rotations, which plotted against the mid-cycle displacements $\mu^{C}$
and $\theta^{C}$ give the expected mid-cycle exponents of $\approx-0.8$
and $\approx-0.6$ as seen on figure \ref{fig:Irreversibility}.

\begin{figure}[H]
\includegraphics[width=0.5\columnwidth]{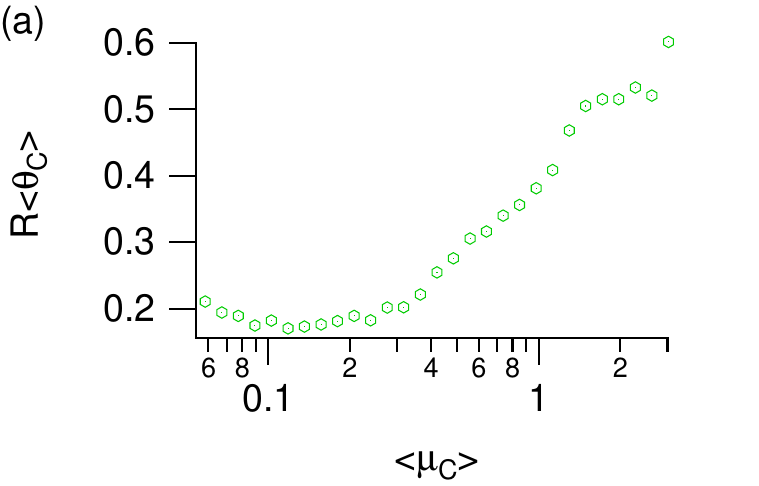}\includegraphics[width=0.5\columnwidth]{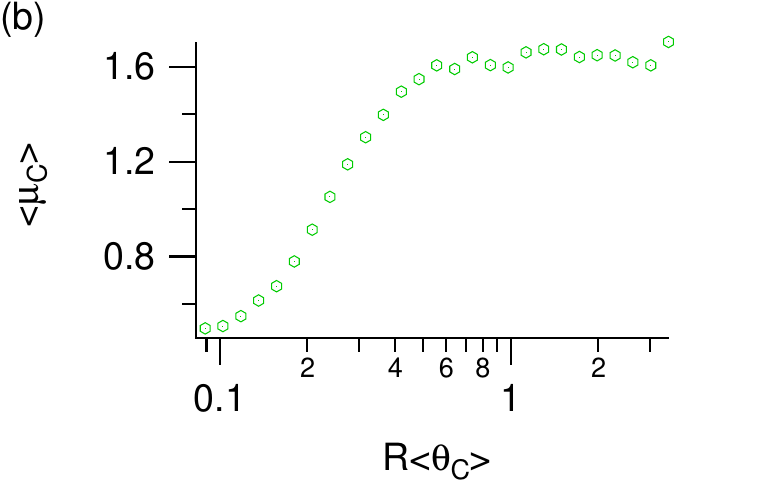}

\includegraphics[width=0.5\columnwidth]{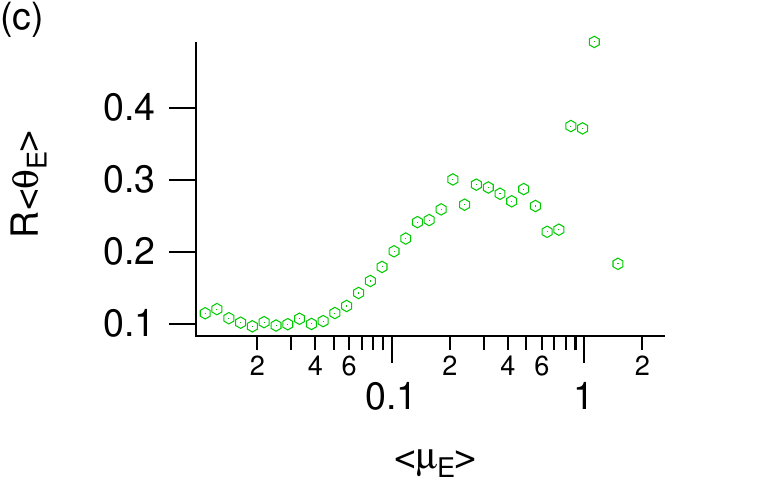}\includegraphics[width=0.5\columnwidth]{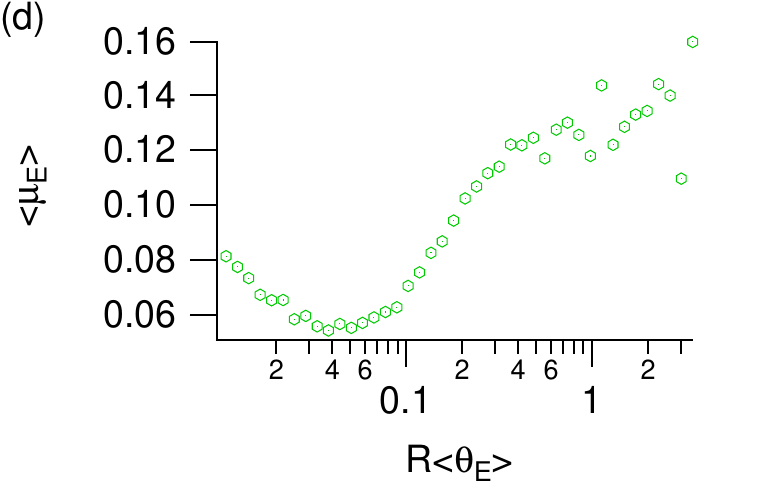}

\caption{\label{fig:Correlations}(a) Averaged value of rotational displacement
binned over values of translational displacement at time $C$. (b)
Averaged value of translational displacement binned over values of
rotational displacement at state C. (c) and (d) same as (a) and (b)
at time \emph{$E$.} All units in mm.}
\end{figure}

We have found that the expected middle or end-of-cycle rotational
displacements $\theta^{C,E}$ and translational displacements $\mu^{C,E}$
are very weakly correlated, as can be seen on figure \ref{fig:Correlations}.
This result is in contrast to the case of ellipsoidal particles, where
a strong correlation between rotations and translations was experimentally
found \cite{KOU:2018:ID757}. The difference can be explained by the
fact that for ellipsoidal particles, rotations imply a change in the
effective volume occupied by the particles and thus can couple to
translations of surrounding particles.

\begin{figure}
\includegraphics[width=1\columnwidth]{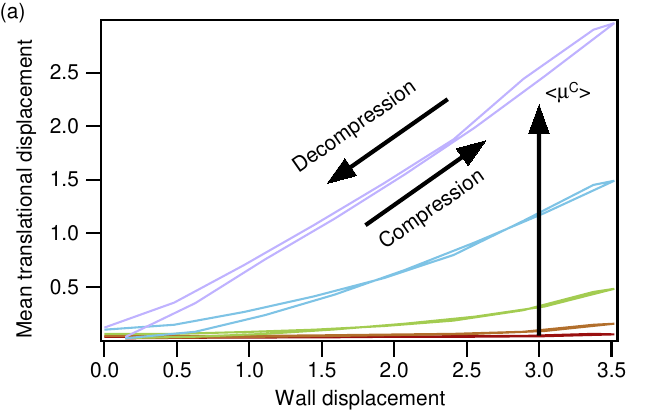}

\includegraphics[width=1\columnwidth]{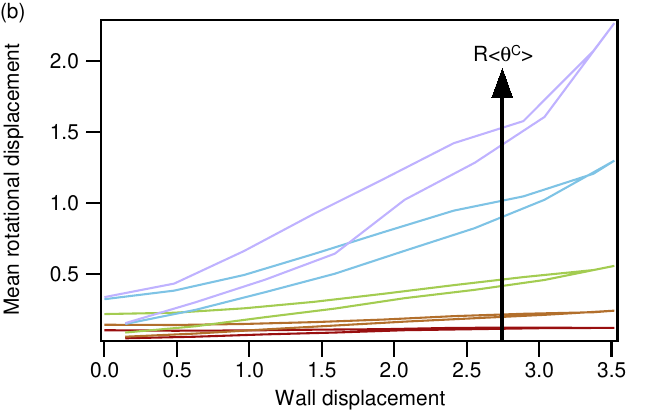}
\caption{\label{fig:Cycles of motion}Mean translational displacement $\left\langle \mu^{T}\right\rangle $
trajectories (a) and rotational displacement $R\left\langle \theta^{T}\right\rangle $
trajectories (b). To compute this trajectories we average over particles
with comparable mid-cycle displacements represented by different colors.
All units in mm. The mid-cycle displacements go from $\mu^{C}=6\times10^{-2}\thinspace mm$
to $3\thinspace mm$ for translations and from $R\theta^{C}=1\times10^{-1}\thinspace mm$
to $3\thinspace mm$ for rotations. Results for 2.5\% compression
amplitude.}
\end{figure}

Beyond the reversibility to its initial position and orientation,
we investigate the reversibility of the whole particle trajectory,
both in terms of translation and rotation. We plot the displacements
of particles during a cycle of compression averaged over similar values
of mid-cycle displacements (Fig. \ref{fig:Cycles of motion}). These
results are based on the actual motion of the moving wall, corrected
for slight backlash present in our experimental system.

Figure \ref{fig:Cycles of motion} highlights the difference between
the behavior of granular translations and rotations. Translations
are very reversible, with beads following much of the same path during
compression and dilation, especially at high values of the wall displacement.
On the other hand, the rotation trajectories are much less so; the
dilation path, which begins at maximum wall displacement and goes
back to zero wall displacement, diverges almost immediately from the
compression path. In other words, in contrast to the translations,
the forward and backward paths for rotations differ significantly
even close to point of wall reversal (maximum wall displacement).
The reversibility of both translations and rotations improves after
many cycles as compared with the beginning of the experiment (not
shown), suggesting that, after an initial transient period, the system
self-organizes into a more reversible configuration.

\section{Contact dynamics\label{sec:Contact-dynamics}}

For particles to revert back to their original positions relative
to their neighbors without reversal of orientations requires some
change in frictional contacts. To study the contacts dynamics, we
define particles that are within a cutoff distance $r_{c}$ of each
other as being likely in contact. To capture all likely contacts we
chose $r_{c}$ as the distance within which particles have on average
6 contacts per particle. We have verified that varying $r_{c}$ such
that the mean number of contacts per particles is between 4, the isostatic
limit for frictional particles \cite{HANIFPOUR:2015:ID747}, and 6,
the jamming point, does not change the quantitative results presented
here.

\begin{figure}
\includegraphics[width=1\columnwidth]{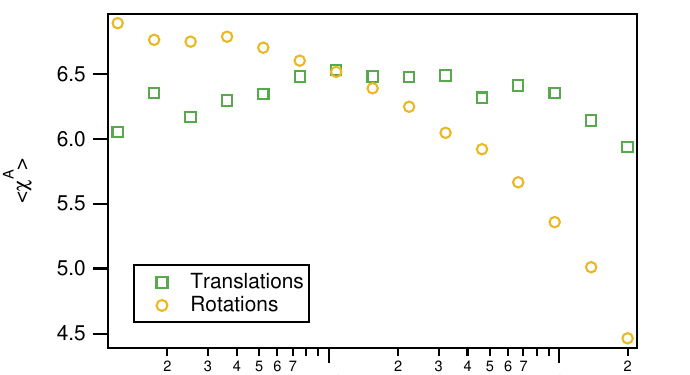}

\caption{\label{fig:contacts_displacement}The mean number of contacts at the
beginning of cycle as a function of translational and rotational displacements.
Displacements in mm. Results for 2.5\% compression amplitude.}
\end{figure}

We define $\chi_{i}^{T}$ as the total number of contacts of an individual
particle at time $T$. While the translational displacement are weakly
correlated with the number of contacts a particle have at the beginning
of the cycle, the rotational displacement are strongly dependent on
this quantity as shown on Figure \ref{fig:contacts_displacement}.
Particles with fever contacts tend to have much bigger end of cycle
rotations. This suggests that particles that can rotate more freely,
are less prone to return to their original positions. Note that in
frictional contacts the transition between the ``rolling'' and ``slipping''
phases is hysteretical. This means than even a conservancy of contact
points does not guarantee the reversal of rotational motion. Indeed,
while we find that both translational and rotational irreversibility
are strongly correlated with a gain or loss of a contact (not shown),
translations are more affected.

\section{Summary and conclusions}

We conducted a novel experiment allowing us to track translations
and rotations of particles in three dimensions during cycles of uni-axial
compression. We found that the overall reversibility of translations
is much higher than that of rotations. This is explained by the absence
of direct correlations between the rotations and translations; the
reversible motion of particles ``trapped'' in cages \cite{ROYER:2015:ID759}
does not imply the reversibility of rotations. Given the available
results of a presence of correlations between translations and rotations
for ellipsoid particles \cite{KOU:2018:ID757}, a larger correlation
between translations and rotations may be observed for larger friction
and different frictional properties of the particles. In our system,
the non-reversibility of rotations is a possible mechanism for the
system to maintain a memory of cycling, even under small forcing conditions
that do not imprint a memory in \textquotedbl{}bulk\textquotedbl{}
indicators of prior forcing (segregation, compaction).

Even though most particles retain their neighbors in our experiment,
their contact-point dynamics follows a more hysteretic path and rotational
motion is not reversible. It is highly probable that the irreversibility
of rotations is also reflected in a parameter which we cannot track
in our experiments: the force distributions. To elucidate this hypothesis,
further work is needed to track the rotations and the forces on the
particles at the same time, either through simulations or different
experimental methods \cite{HURLEY:2016:ID760}. Another important
aspect of granular flow is the vector of rotation compared to the
direction of shear/compression. It has been predicted \cite{HALSEY:2009:ID745}
that particles will rotate perpendicular to the shearing motion. However,
to verify this hypothesis experimentally, we need to access all three
degrees of rotations, which will be the subject of our future work. 

Note that measurements of rotations are particularly valuable in dense
systems such as ours, where translational motion is suppressed and
the main mechanism of energy dissipation is by friction between the
particles. Frictional dissipation manifests itself in rotational motion,
not translation, and thus particle rotations are the key to gain insights
into the energetics of dense granular flow. We expect that the experimental
data presented here will allow for more direct validation of the competing
approaches toward simulating frictional dissipation \cite{BAGI:2004:ID864,WANG:2015:ID865}. 
\begin{acknowledgments}
We are grateful to Zackery Benson, Jacob Prinz, Charlotte Slaughter
and Dara Storer for their help with the experiments. This work was
supported by National Science Foundation grant DMR5244620. C. S. and
D. S. were supported by TREND, an NSF REU program PHY1756179. 
\end{acknowledgments}

\bibliographystyle{aipnum4-1}
\bibliography{references_database}

\end{document}

\title{Supplemental material for \\
 Reversibility of granular rotations and translations}

\author{Anton Peshkov}

\affiliation{IREAP, University of Maryland, College Park, Maryland, USA}

\author{Michelle Girvan}

\affiliation{Departments of Physics, IPST and IREAP, University of Maryland, College
Park, Maryland, USA}

\author{Derek C. Richardson}

\affiliation{Department of Astronomy, University of Maryland, College Park, Maryland,
USA}

\author{Wolfgang Losert}

\affiliation{Departments of Physics, IPST and IREAP, University of Maryland, College
Park, Maryland, USA}
\maketitle

\section{Irreversibility of displacements}

\begin{figure}[H]
\includegraphics[width=0.5\textwidth]{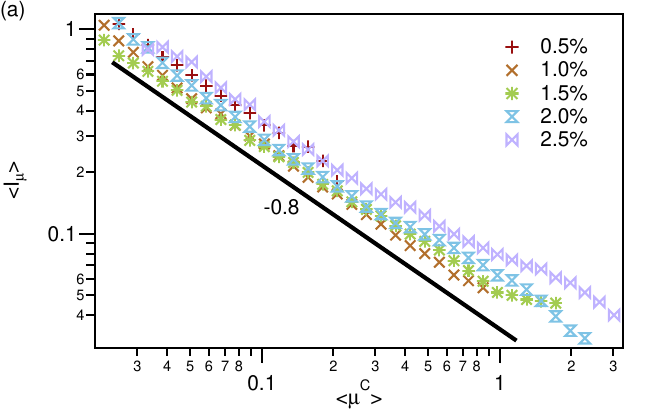}\includegraphics[width=0.5\textwidth]{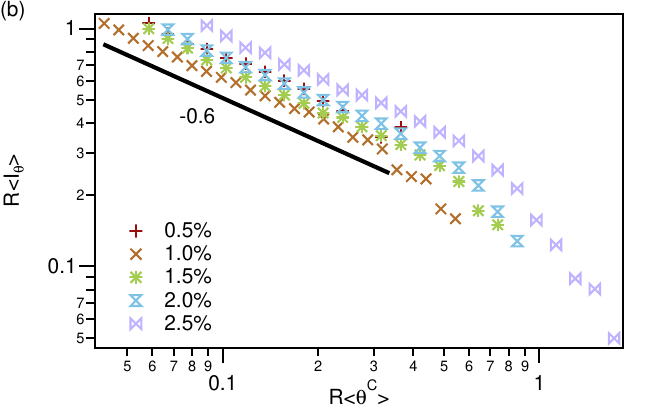}

\caption{\label{fig:Irreversibility}Expected relative irreversibility parameters
$I_{\mu}$ (a) and $I_{\theta}$(b) as a function of the mid-cycle
mobility $\mu^{C}$ and $\theta^{C}$. Displacement amplitudes are
in mm.}
\end{figure}

In addition to studying the translational and rotational displacements,
we can introduce the relative irreversibility parameters $I_{\mu,i}=\nicefrac{\mu_{i}^{E}}{\mu_{i}^{C}}$
for translations and $I_{\theta,i}=\nicefrac{\theta_{i}^{E}}{\theta_{i}^{C}}$for
rotations. As seen in figure \ref{fig:Irreversibility}, both of these
show decaying power laws as a function of the mid-cycle mobility,
with exponents of $\approx-0.8$ and $\approx-0.6$ for translations
and rotations respectively.

\section{Correlations between translations and rotations}

\begin{figure}[H]
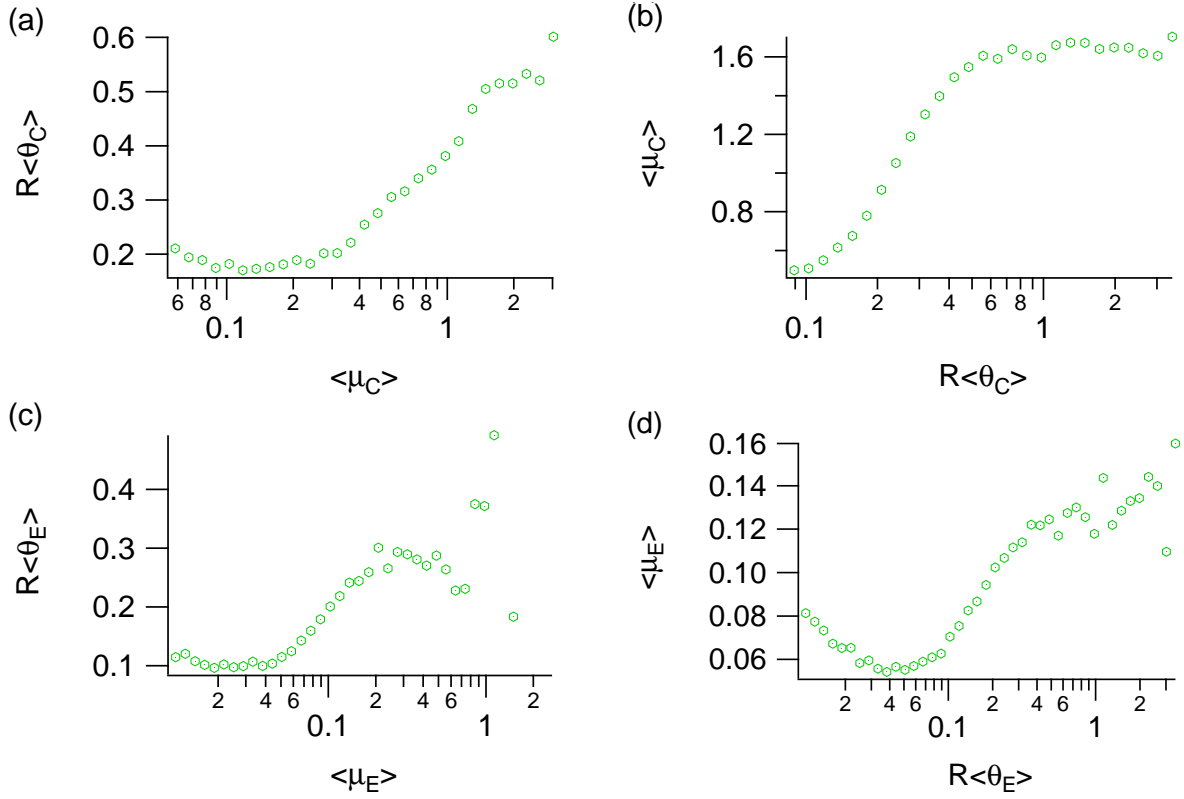

\includegraphics[width=0.5\textwidth]{\string"Images/rot_disp_hc_corr\string".pdf}\includegraphics[width=0.5\textwidth]{\string"Images/disp_rot_hc_corr\string".pdf}

\includegraphics[width=0.5\textwidth]{\string"Images/rot_disp_fc_corr\string".pdf}\includegraphics[width=0.5\textwidth]{\string"Images/disp_rot_fc_corr\string".pdf}

\caption{\label{fig:Correlations}(a) Averaged value of rotational displacement
binned over values of translational displacement at state C. (b) Averaged
value of translational displacement binned over values of rotational
displacement at state C. (c) and (d) same as (a) and (b) at state
E\emph{.} All units in mm.}
\end{figure}

To investigate the presence of correlations between rotations and
translations, we plot averaged values of one against the other in
figure \ref{fig:Correlations}. While small variations are present
they are negligible compared to the decades of variation of mobilities
seen in figure 3 of the main article.

\section{Cycles of displacement}

In addition to plotting the cycles of translational and rotational
displacements at the end of the experiments we can plot the same quantities
at the beginning of the experiment to compare the change in the irreversibility.

\begin{figure}[H]
\includegraphics[width=0.5\textwidth]{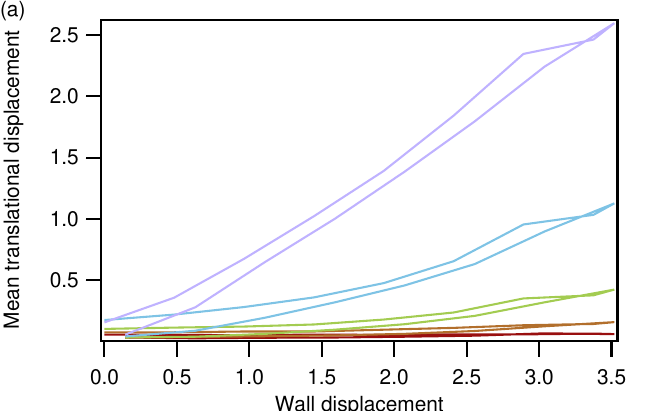}\includegraphics[width=0.5\textwidth]{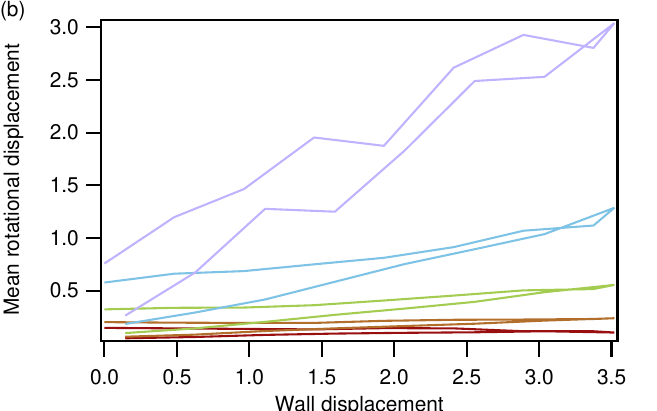}

\caption{\label{fig:Cycles}Bead translational (a) and rotational (b) displacement
cycles averaged over identical mean mobilities for the first five
cycles. Compare with figure 4 in the article where the same quantities
are plotted for the last ten cycles of the experiment. The mid-cycle
mobility goes from $6\thinspace10^{-1}\thinspace mm$ to $3\thinspace mm$
for translations and from $1\thinspace10^{-1}\thinspace mm$ to $3\thinspace mm$
for rotations. All units in mm.}
\end{figure}